\documentclass{emulateapj}

\usepackage{graphicx}
\usepackage{subfigure}
\usepackage{latexsym}

\newcommand{\HI}{\protect\ion{H}{1}}

\newcommand{\msun}{$M_\odot$}
\newcommand{\etal}{{et~al.}}
\newcommand{\cmsq}{cm$^{-2}$}
\newcommand{\mhi}{$M_{HI}$}

\newcommand{\kms}{~km\,s$^{-1}$}

\newcommand{\vgsr}{$v_{\rm GSR}$}

\defcitealias{pis04}{Paper~I}
\defcitealias{pis07}{Paper~III}

\slugcomment{To appear in ApJ July 1, 2007, vol. 663, num. 1}

\begin{document}

\title{An \HI\ survey of six Local Group analogs:  I. Survey description and 
the search for high-velocity clouds.}
\author{D.J. Pisano\altaffilmark{1,2}}
\affil{Naval Research Laboratory, Code 7213, 4555 Overlook Ave. SW, 
Washington, D.C., 22314, USA}
\email{dpisano@nrao.edu}
\author{David G. Barnes}
\affil{Centre for Astrophysics \& Supercomputing, Swinburne University, 
Hawthorn, Victoria 3122, Australia}
\email{David.G.Barnes@gmail.com}
\author{Brad K. Gibson}
\affil{Centre for Astrophysics, University of Central Lancashire, Preston, 
PR1 2HE, UK}
\email{bkgibson@uclan.ac.uk}
\author{Lister Staveley-Smith}
\affil{School of Physics M013, University of Western Australia, Crawley, WA 
6009, Australia}
\email{Lister.Staveley-Smith@uwa.edu.au}
\author{Ken C. Freeman}
\affil{RSAA, Mount Stromlo Observatory, Cotter Road, Weston, ACT 2611, 
Australia}
\email{kcf@mso.anu.edu.au}
\author{Virginia A. Kilborn}
\affil{Centre for Astrophysics \& Supercomputing, Swinburne University, 
Hawthorn, Victoria 3122, Australia}
\email{vkilborn@astro.swin.edu.au}

\altaffiltext{1}{National Research Council Research Fellow}
\altaffiltext{2}{Current Address: NRAO, P.O. Box 2, Green Bank, WV 24944}

\begin{abstract}

We have conducted an \HI\ 21 cm emission-line survey using the
Parkes 20cm multibeam instrument and the Australia Telescope Compact Array
(ATCA) of six loose groups of galaxies chosen to be analogs to the Local
Group.  The goal of this survey is to make a census of the \HI-rich
galaxies and high-velocity clouds (HVCs) within these groups and compare
these populations with those in the Local Group.  The Parkes observations 
covered the entire volume of each group with a rms \mhi\ sensitivity of 
4-10$\times$10$^5$\msun\ per 3.3 \kms\ channel.  All potential sources 
detected in the Parkes data were confirmed with ATCA observations at 
$\sim$2\arcmin\ resolution and the same \mhi\ sensitivity.  All the confirmed 
sources have associated stellar counterparts; no starless \HI\
clouds--HVC analogs--were found in the six groups.  In this paper, we
present a description of the survey parameters, its sensitivity and
completeness.  Using the population of compact HVCs (CHVCs) around the
Milky Way as a template coupled with the detailed knowledge of our
survey parameters, we infer that our non-detection of CHVC analogs implies 
that, if similar populations exist in the six groups studied, the CHVCs 
must be clustered within 90 kpc of group galaxies, with average 
\mhi$\lesssim 4\times$10$^5$\msun\ at the 95\% confidence level.  The 
corollary is that the same must apply to Milky Way CHVCs.  
This is consistent with our previous results from a smaller sample of 
groups, and in accordance with recent observational and theoretical 
constraints from other authors.  These results confirm that there is very 
little neutral matter around galaxies, and that any substantial reservoir of 
baryons must be in other phases.  
\end{abstract}


\section{Introduction}
\label{intro}

The majority of galaxies reside in groups \citep{tul87}.  Loose groups of 
galaxies, such as the Local Group in which the Milky Way is located, are a 
collection of a few large, bright galaxies and tens of smaller,
fainter ones.  The large galaxies are typically separated by a few
hundred kiloparsecs and spread over an extent of approximately a
megaparsec. Groups of galaxies are the building blocks of galaxy clusters, 
feeding gas-rich spiral galaxies into the cluster
\citep[e.g.][]{bur94,bra00,gon05}.  Loose groups themselves are  still
collapsing and not virialized \citep{zab98}.  The gas-rich
galaxies within loose groups are also still forming as they accrete
their dwarf galaxy satellites \citep[see][and references therein]{fre02} 
and their reservoir of gas \citep[e.g.][]{bli99}.  Some of this reservoir 
may be in the form of neutral gas and possibly related to the  
``high-velocity clouds'' (HVCs) seen around the Milky Way 
\citep[see][and references therein for a review]{wak97,vanW04}.

HVCs are clouds of neutral hydrogen (\HI) discovered approximately 40
years ago \citep{mul63} covering up to 37\% of the sky \citep{mur95, loc02}
with velocities inconsistent with simple Galactic rotation and in
excess of $\pm$90 \kms\ of the Local Standard of Rest \citep{wak97}.
They lack associated stellar emission \citep{sim02,wil02,sie05}, and
so we have no direct measure of their distances, and, therefore, their
masses.  As such, we can not easily discriminate between their
possible origins.  Nevertheless, HVCs likely represent a variety of
phenomena.

Some HVCs may be related to a galactic fountain \citep{sha76,bre80}.
Other HVCs are certainly tidal in  origin:  the Magellanic Stream is the 
most obvious of these features, formed  via the tidal interactions between
the Milky Way, Large Magellanic Cloud, and  Small Magellanic Cloud
\citep[e.g.][]{put02}. Other HVCs may be related to the
Sagittarius dwarf \citep{put04}.  Some HVCs may even be satellites
unto themselves \citep{loc03}.  \citet{oor66,oor70} originally
proposed that HVCs may be infalling primordial gas associated with the
formation of the Milky Way.  Complex C may be such an example
\citep{wak99,tri03,gib01}

While the idea of associating HVCs with galaxy formation is not new,
recently this hypothesis has attracted more attention.  \citet{bli99}
and \citet{bra99} suggested that the HVCs and the compact HVCs
(CHVCs),  respectively, may contain dark matter and could be related
to the small  dark matter halos predicted to exist in large numbers by
cold dark matter  models of galaxy formation
\citep[e.g.][]{kly99,moo99}.  As originally conceived by
\citet{bli99} and \citet{bra99}, HVCs or the subset of the most compact HVCs
(CHVCs) have distances $\sim$1 Mpc, and \mhi $\sim$10$^7$\msun. 
Current models associating CHVCs with dark matter
halos by \citet{deh02b} suggests the CHVC distribution  has a Gaussian
distribution about the  Milky Way and M31 with  D$\sim$150-200 kpc and
\mhi $\sim$10$^{5.5-7}$\msun.   A similar distance distribution is
suggested by \citet{mal04} who propose that the HVCs are the products
of the cooling, condensing hot halo.  Such clouds, however, would not
be associated with dark matter halos.

There have been many detections of possible HVCs around other
galaxies.  In these systems, we can easily determine the masses and 
separation of the clouds.  We can clearly see examples of extragalactic 
\HI\ clouds associated with galactic fountains, tidal debris, and even very 
faint dwarf galaxies, but none to date can be unambiguously identified as
being associated with galaxy formation or dark matter halos.  The \HI\
Rogues Gallery \citep{hib01} \footnote{available online at 
\url{http://www.nrao.edu/astrores/HIrogues/}} contains many examples
of such \HI\ clouds.

\citet{sch94} identified high-velocity wings on integrated  \HI\
profiles from a sample of face-on galaxies.  The presence of these
wings  was well-correlated with an increased star formation rate in
the galaxy, and so it is likely that \citet{sch94} identified \HI\
associated with a galactic fountain.  \citet{kam93} and \citet{boo04}
have identified high-velocity \HI\ associated with NGC 6946.  Much of
this gas is coincident with \HI\ holes and is predominately
coincident with the optical disk.  As NGC 6946 is prolifically forming
stars, and has had eight supernovae in the past century, it
seems likely that most of this gas must arise from a galactic
fountain.   Other galaxies, such as NGC 891 \citep{swa97}, NGC 253
\citep{boo05}, NGC 4559 \citep{bar05}, NGC 2403 \citep{fra02}, also
show signatures of extra-planar \HI\  possibly associated, at least in
part, with a galactic fountain.

We know of many examples of gaseous tidal debris without optical counterparts
around individual galaxies. The Leo Ring is a 200 kpc diameter ring of \HI\ 
orbiting within  the M~96 Group \citep{sch83,sch85}.  While it may have a 
tidal origin, as it appears to be interacting with M~96 \citep{sch85}, its 
orbital period of 4 Gyr and its coherence over such a large area is a
challenge for such an explanation.  Instead, it may be that this ring
is a primordial remnant of the formation of the group \citep{sch85}.
A $\sim$50 kpc diameter ring ($\sim$100 kpc in length), a similar feature to
the Leo Ring but on a much smaller scale, is seen around  NGC 4449
\citep{hun98}.  While this could be explained as ongoing accretion of
primordial gas \citep{hun98}, the structure of this ring (if not
its presence) can be explained via a recent interaction with a nearby
companion \citep{the01}.  IC~10 has an extended \HI\ distribution
that is  counter-rotating with respect to the main body of the galaxy,
plus a long streamer extending away from the galaxy \citep{wil98}.
\citet{wil98} suggest this \HI\ represents ongoing infall of
primordial material, and the absence of a nearby perturber, despite
residing in the Local Group, provides  circumstantial support for this
scenario.  NGC~925 represents our final example of intergalactic \HI\
which may be primordial or tidal in origin.   NGC~925 has a
$\sim$10$^7$\msun\ \HI\ cloud attached to the main galaxy by a
streamer \citep{pis98}.  NGC~925 does reside in a group, but the
nearest galaxy to NGC~925 is 200 kpc away.  Is this \HI\ cloud then
primordial gas, the debris from an old tidal encounter, or a disrupted
dwarf galaxy?

\citet{pis02} searched for \HI\ clouds around 41
quiescent, isolated galaxies and found only gas-rich dwarf galaxies.
The absence of starless \HI\ clouds around isolated galaxies suggests
that the presence of such clouds in denser environments (such as those
described above) may be the result of interactions \citep{pis02}.  
To date, all claims of intergalactic \HI\ clouds without associated stars, 
or ``dark galaxies'' \citep[e.g. \HI\ 1225+0146][]{gio89} have 
turned out to be, on closer inspection, either low surface brightness
galaxies \citep{mcm90} or tidal debris connected with a bright galaxy 
\citep{bek05a,bek05b}.  These detections tend to be in dense environments, 
where tidal interactions are more probable.  As such, if we want to find HVC 
analogs associated with galaxy formation, it is important to search for such 
\HI\ clouds, using very sensitive instruments, in low density environments 
similar to the Local Group. 

There have been many previous surveys of galaxy groups searching
for starless \HI\ clouds.  \citet{lo79}, \citet{kra99}, and \citet{kov05} 
surveyed the entirety of the Canes Venatici I group.  They found many 
new dwarf galaxies, but no HVC analogs.  \citet{zwa00}, \citet{zwa01} and 
\citet{deB02} conducted a sparse survey of nearby groups, again finding no 
HVC analogs.  The HIPASS survey \citep{bar01} covered the 
entire southern sky, and, while some \HI\ clouds were detected 
\citep[e.g.][]{ryd01}, these are likely tidal in origin.  The deepest survey 
of a group to date was the HIDEEP survey of the Cen A group, with a 
5$\sigma$\mhi\ limit of $2\times 10^6$\msun, but all of the new \HI\ 
detections in this survey had optical counterparts
\citep{min03}.  The Cen A group is dominated by a large elliptical galaxy, so
it may not be a good analog to the Local Group, unlike those studied by the
other collaborations, and, hence, perhaps not the ideal location to search
for HVC analogs.  

Our survey, as first reported in \citet[][hereafter Paper~I]{pis04}, seeks to 
provide the deepest observations of the entirety of six loose groups, all 
analogous to the Local Group, to find, or place strong limits, on the presence 
of CHVC analogs around these galaxies.  Paper~I reported the initial results 
of this work for the first three groups.  In this paper (Paper~II), we will 
discuss the survey properties in greater detail, including its sensitivity and
completeness, and present the results of our search for CHVC analogs for all 
six groups  We use our measure of completeness to improve our model from
\citetalias{pis04} and use it to determine the implications of our
non-detections for the distribution of CHVCs in these groups and around the 
Milky Way.  In the final paper of this series \citep[][hereafter 
Paper~III]{pis07}, we will examine the properties of the galaxies in these 
loose groups and how their properties compare to the population of the Local 
Group.  

This paper is structured as follows.  In Section~\ref{sample} we discuss
our sample selection and the basic properties of the groups we observed.  The
Parkes observations and reductions are presented in Section~\ref{obs}, followed
by a discussion of source-finding in Section~\ref{sources}.  The confirming 
interferometric observations and reductions are discussed in 
Section~\ref{atca}.  In order to determine the implications of our 
failure to detect any HVC analogs, we must have a good understanding of the 
completeness of this survey; this is presented in Section~\ref{comp}.  The
results of our survey are briefly summarized in Section~\ref{results}.  The 
implications for the distribution of CHVC analogs and the nature of the 
HVCs around the Milky Way are discussed in Section~\ref{disc}, and we
conclude in Section~\ref{conc}.  

\section{Sample Selection}
\label{sample}

We selected nearby (\vgsr\ $<$ 1000 \kms) loose groups of galaxies that
are analogs to the Local Group, which contain only spiral and irregular 
galaxies.  For these nearby groups, we can obtain high spatial resolution
and can detect very low-mass \HI\ clouds.  Groups that may be confused with
Galactic or Local Group \HI\ emission (\vgsr\ $<$ 300 \kms) were avoided.  
In the ``Lyon Groups of Galaxies'' catalog  \citep[LGG;][]{gar93} eight groups
in the Parkes declination range ($\delta < 0$\arcdeg) matched these criteria.  
We chose to observe five groups of galaxies
from this catalog:  LGG 93, LGG 106, LGG 180, LGG 293 and LGG 478.  In
addition, we chose a sixth group with similar properties from the
group  catalog of \citet{ste05}, which is solely composed of HICAT
galaxies \citep{mey04} that we call ``the HIPASS Group''.  Properties  of
these groups are listed in Table~\ref{tab:grpprop} and discussed
below.   We assume an H$_0$ = 72 \kms\ Mpc$^{-1}$ \citep{spe03}, and 
calculated the distances to the groups after
correcting their heliocentric velocities for a multi-attractor
velocity flow model of the local universe \citep[][ Masters \etal\ 2007, in
preparation]{mas05}.  Note that as a result of this new method, the distances 
used here are different from those in \citetalias{pis04}.  The uncertainties 
on these distances are $\sim$2 Mpc (Masters, 2006, priv. comm.).  

\section{Parkes Observations and Reductions}
\label{obs}

\subsection{Multibeam Observations}

The six groups were observed, in sessions of approximately 10 nights each, 
between October 2001 and June 2003 using the 20cm multibeam 
instrument\citep{sta96} on the Parkes 64m radiotelescope.  The multibeam has 
13 hexagonally-packed receivers. Each individual half-power beamwidth is
14.3$\arcmin$ in diameter and is separated by two beamwidths from its
neighbor.  The whole system spans an on-sky diameter of 1.7$^\circ$.   
For LGG 93 and LGG 180, we observed with an 8 MHz bandwidth
using only the inner seven beams.  The velocity resolution for these 
observations was 1.65 \kms.  By August 2002, new 16 MHz filters were 
available for all 13 beams, and were used for the observations of the 
remaining four groups.  The 16 MHz bandwidth observations had a velocity 
resolution of 3.3 \kms.  Only data taken at night was used to avoid solar
interference which can both add noise to the data cubes and result in
spurious detections.

All groups were observed by scanning the multibeam across the group along 
great circles in both the right ascension and declination directions at a 
rate of 0.5 degrees per minute.  With data being recorded every 5 seconds, 
each beam area is sampled almost six times in the scan direction.  For 
observations with the inner seven beams, we rotated the multibeam by 
19.1$^\circ$ such that the half power points of each beam touch.  For
observations using all 13 beams, the rotation angle was 15$^\circ$ to
approximate the same coverage.  Each scan was offset by 13$\arcmin$ so that 
the field is uniformly sampled perpendicular to the scan direction by each 
beam.  We cycled through all the scans covering the full area of
the group,  in both directions, until the desired sensitivity was
reached.  The  characteristics of our observations for each group are
listed in  Table~\ref{tab:obsprop}.

\subsection{Calibration}

A default calibration based on past measurements is
applied to the data as it is written to disk.  This default
calibration was determined for the HIPASS setup with 64 MHz bandpass
and a central frequency of 1384 MHz.  We regularly (usually once a day) 
observed Hydra~A to check the calibration using our observing 
configuration, and used our derived average calibration factors instead 
of the defaults in our reduction process.  The derived factors were typically 
within 10\% of the default values.  

\subsection{Reductions}

All data were reduced and gridded using the {\sc livedata} and {\sc
gridzilla}  packages in {\bf{\sc aips++}}.  The details of how these
packages work and were used for reducing HIPASS data are discussed 
in \citet{bar01}.  We used a two-step iterative process in reducing 
our data.

{\sc livedata} was used to remove the bandpass both temporally and
spectrally.  The bandpass was fit in the scan direction with a first
order polynomial, where outlying points (from the source or
interference) were clipped at the 1.5 $\sigma$ level over  the
successive iterations.  This is equivalent to subtracting and
normalizing by an  `off' position for a standard `on-off' observation.
A first order spectral baseline was then fit to the resulting
spectrum at each position of each individual scan.  Like for the
HIPASS reductions, a 25\% Tukey smoothing function was applied to damp
out the ringing resulting from strong line sources (like the Galaxy).
The reduced scans were then combined into a single data cube using the
program {\sc gridzilla}.  {\sc gridzilla}  determines which spectra
contribute to which pixel and with what weight they contribute,  while
discarding outlying spectra.  We used a mean weighting scheme while
clipping the brightest 2\% of the spectra.  Since there were a few hundred
spectra contributing to a given pixel, this resulted in a very small
loss in signal-to-noise, while removing most of the effects of
interference.  We were able to use mean griding as compared to the
median griding used for HIPASS because our data were all taken at
night eliminating contamination from solar interference, but some other
forms of interference are visible in our data cubes.  This also results in
an improvement in the noise.  

At this point, the cubes were searched for sources (more on this in
Section~\ref{sources}).  All sources were then masked, and the
reduction process was repeated with a few small differences.  In {\sc
livedata}, a second order polynomial was fit in the time (spatial)
direction with 2$\sigma$ clipping over three iterations.  A second
order polynomial was also fit and removed from the spectral baselines.
Because the sources are masked, these higher order fits yield a
flatter residual baseline while lacking large negative sidelobes
around bright sources that can hide nearby weak sources. Furthermore, this
process yields more robust measures of the \HI\ spectrum of all
sources.  The resulting noise level and mass sensitivities are listed
in Table~\ref{tab:obsprop}.

\section{Source Finding}
\label{sources}

Once the data cubes for each group were complete, three teams of authors 
(DJP, DGB, and BKG \& VAK in tandem) searched each cube by
eye for detections.   Any source identified in two or more searches
was considered ``real'' and was  a candidate for follow-up
confirmation with the Australia Telescope Compact Array\footnote{The
Australia Telescope Compact Array (ATCA) is part of  the Australia
Telescope, which is funded by the Commonwealth of Australia for
operation as a National Facility managed by CSIRO.}.  Each cube had simulated
sources inserted into the data by a third party (Dr. M Zwaan) to provide a 
measure of the 
completeness of our survey as a function of linewidth and integrated flux.  
This process was done in the same fashion as for HIPASS \citep{zwa04}.  
As described above, after all the possible detections were identified, the 
final, masked cubes were made and re-searched to identify any new 
sources--none were found.  

\section{ATCA Observations and Reductions}
\label{atca}

We used the ATCA to observe all 105 of the 112 Parkes detections in and behind 
the six groups to confirm the reality of all our detections, to uniquely 
identify optical counterparts, and resolve any confused sources.  The 
remaining seven galaxies were behind one of the groups and previously detected 
in HIPASS.  Of the remaining 105 possible sources, 15 had previously been 
observed with the ATCA or VLA. We observed the remaining 90 between October 
2002 and March 2005 using a compact, 750m configuration for 
observations of all of the groups to obtain a nearly circular beam of 
$\sim$1-2$\arcmin$.  Because of the near equatorial position of LGG 293, it was
the exception; we observed its galaxies with the H214C configuration 
(a configuration with antennas on the ATCA's North Spur with a maximum 
baseline of 214m) for resulting resolution of $\sim$2-3$\arcmin$.  We used a 
bandwidth of 8~MHz ($\sim$1700 \kms)
with 512 channels of 3.3 \kms\ after Hanning smoothing.  We observed
1934-638 as a primary flux and bandpass calibrator at the beginning
or end of each day's observing run, and made hourly observations of
fainter, closer radio sources for phase calibration.  To confirm all
our detections, every source was observed with the ATCA to at least the same
sensitivity as in the Parkes data, typically about 4 mJy/beam.
The majority of the observations occurred at night to minimize the effects 
of solar interference.

The calibration and reduction of the data was done in the standard way
using MIRIAD.  Data were flagged by hand to remove interference, both
terrestrial and solar.  Generally very few data needed to be
flagged.  The data for each day were separately reduced and
calibrated.  The line-free region in each cube was identified and the
data were continuum subtracted.  The line data from each night were
then combined to make a data cube using a robust (a robustness
of two) or natural weighting scheme.
The cubes were then CLEANed down to $2-3$ times the RMS noise level,
whichever was sufficient to remove all CLEAN artifacts.  Moment maps of 
the total intensity and velocity field (moments 0 and 1) were 
made in AIPS with $2-3\sigma$ blanking over the velocity range where 
emission was observed.   

As our groups are nearby and composed of large, bright galaxies, many
of the galaxies in our groups had been previously observed with the
VLA and the ATCA.  We have taken these data from the relevant on-line
archives and re-reduced the data in the same manner described above.  
This archival data generally had a better sensitivity than our Parkes
data, with velocity and spatial resolution comparable to our ATCA 
750m array observations.  

All cubes were inspected by eye for \HI\ emission from the Parkes
detection and for additional emission from HVC analogs or gas-rich dwarf
galaxies.  Of the original 112 Parkes detections, only five were not
confirmed with the ATCA or VLA.  An additional four dwarf galaxies,
behind the target groups, were resolved in the interferometric data.
No HVC analogs were detected in the ATCA or VLA data,
despite the higher spatial resolution and equivalent mass sensitivity
as compared to the Parkes data.  All of these data will be presented in more 
detail in \citetalias{pis07}. 

\section{Completeness}
\label{comp}

In order to obtain an accurate census of the \HI-rich objects in the six
loose groups, it is essential to have a measure of the completeness 
of our survey, particularly the completeness as a function of linewidth, 
$W_{20}$ and integrated flux, $S_{int}$, as we derive most properties 
of the galaxies from these two quantities.  In \citetalias{pis04}, we made
the blanket assumption that our survey was 100\% complete for sources with
an integrated flux brighter than 10$\sigma$, and 100\% incomplete for
fainter sources.  In this paper, we refine our measures of completeness. 

The completeness was evaluated in three ways.  First, we examine the 
distribution of the number of sources (real and simulated) as a function of 
$W_{20}$ and $S_{int}$ independently.  Second, we examine the distribution of 
simulated sources as a joint function of $W_{20}$ and $S_{int}$.  Finally, 
we compare our data with the completeness function for HIPASS.  All methods 
have their advantages and drawbacks.

\begin{figure}[hb]
\begin{center}
\includegraphics[scale=0.7, angle=-90]{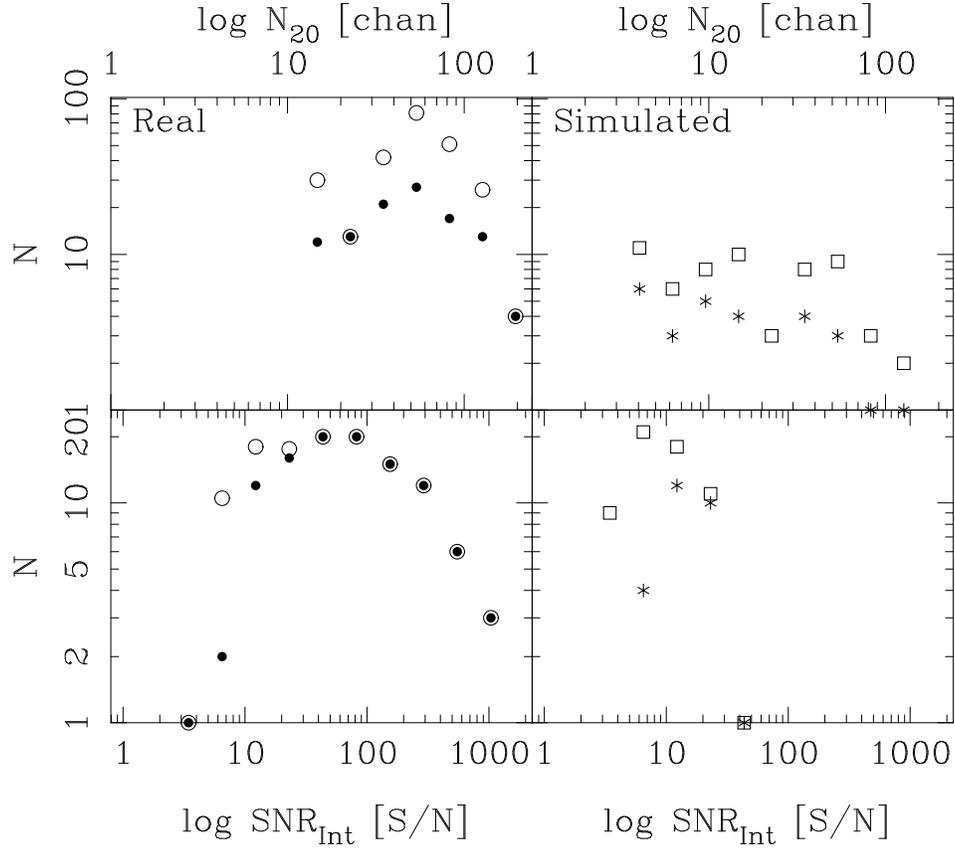}
\caption{The number of detected sources as a function of 
$N_{20}$ (top) and $SNR_{int}$ (bottom) in units of channels and theoretical 
integrated signal-to-noise ratio, respectively.  The left two panels show the
real sources as filled circles.  The right two panels are for the simulated 
sources, with the totals (open squares) and those detected (stars).  The open
circles in the left panels reflect the real data corrected for incompleteness
based on the simulated sources in the right panels.  No correction is applied
if no simulated sources were detected.
\label{fig:sw_comp}}
\end{center}
\end{figure}

Figure~\ref{fig:sw_comp} shows the number of real and simulated sources 
(both detected and undetected) as a function of $W_{20}$ and $S_{int}$.  
Because this plot represents the composite completeness
for six fields with slightly different sensitivities and channel
sizes, we must find a way to scale the parameters to a common scale.  
We use the following formulae to do so:

\begin{equation}
N_{20} = \frac{W_{20}}{\Delta V}
\label{eq:w}
\end{equation}
\begin{equation}
SNR_{int} = \frac{S_{int}}{\sigma \Delta V \sqrt{N_{20}}}
\label{eq:s}
\end{equation}

where $\Delta V$ represents the channel size in \kms\ and $\sigma$
represents the noise in a single channel.  $N_{20}$ and
$SNR_{int}$ are in units of channels and integrated
signal-to-noise ratio as compared to the theoretical limit if the
noise is Gaussian.  The completeness is just the ratio of the number
of detected simulated sources to the total number, and this has been
applied to the data in the left panels (as shown by the open circles).  
It is evident that the drop in the number of real sources as a function 
of $SNR_{int}$ is purely due to incompleteness and does not reflect 
the true flux distribution of our sources.  

For $N_{20}$ the completeness is much more uniform, as seen from the
simulated sources.  In fact, the incompleteness appears to have little 
dependence on the velocity width of the source.  The drop in the number of
real sources as a function of velocity width is, again, not just due to 
incompleteness.  We could have smoothed the data and searched
the cubes over a range of velocity resolutions.  In principle, this would
make it easier to detect sources whose velocity widths best match our 
resolution elements.  Since we did all our searching by eye, we chose not
to do this due to the labor involved.  

\begin{figure}
\begin{center}
\includegraphics[scale=0.7, angle=-90]{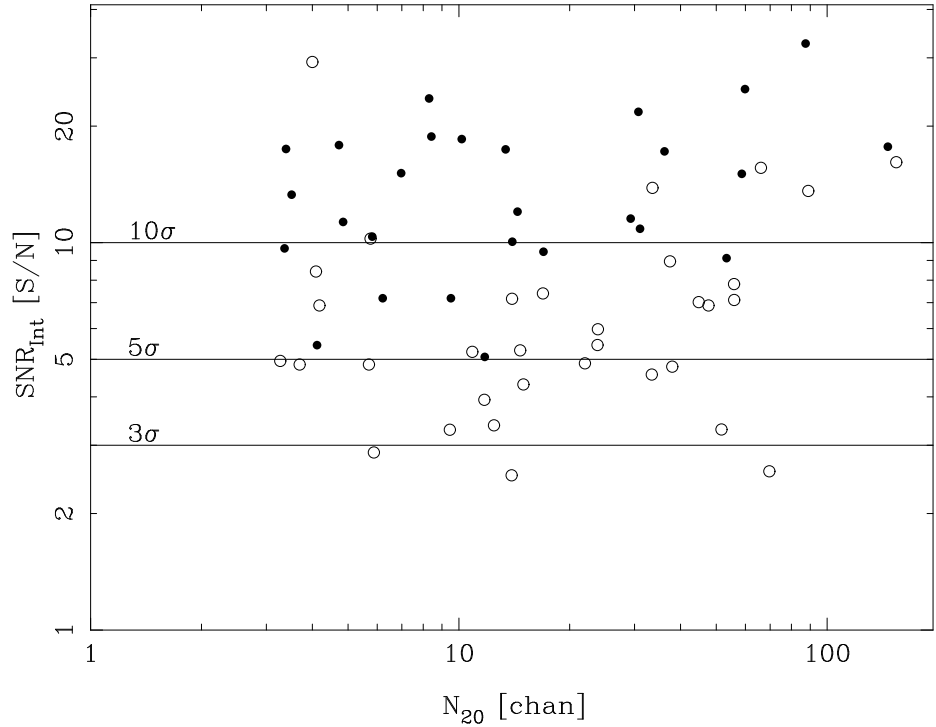}
\caption{The simulated sources that were detected (solid circles) and not 
detected (open circles) as a function of integrated flux (in units of 
integrated signal-to-noise ratio) and linewidth (in units of channels) in all 
six groups.  The flux 
units are defined as the noise in a single channel times the square root of 
the number of channels.  For reference, signal-to-noise ratios of 3, 5, and 
10 are marked on the figure. \label{fig:fakes}}
\end{center}
\end{figure}

Figure~\ref{fig:fakes} shows the properties of all simulated sources, and
those which were identified by more than two authors.  As there were
only 10 simulated sources per cube, we have used Equations~\ref{eq:w}
and \ref{eq:s} to place all the sources on a single plot.  It is evident from
Figure~\ref{fig:fakes} that sources fainter than 5$\sigma$ are not detected 
regardless of their linewidth.
Above this flux level, the presence of ripples in the spectrum, caused by
standing waves from broadband interference can degrade
our ability to detect broad sources.  Even for sources brighter than
10$\sigma$ we are only 76\% complete.  This is due to broad line sources 
with low peak fluxes, only two of ten sources with linewidths smaller than ten
channels and fluxes above 10$\sigma\ $ are undetected, but is also the result 
of small number statistics.  Note also that we have no simulated sources with 
linewidths smaller than three channels.
Such sources would be indistinguishable from narrow band radio
interference, and, as such, highly unreliable.   While we could use
this figure directly to correct our survey for incompleteness, there
are relatively few simulated sources here (only 60) which makes it
difficult to accurately assess the completeness throughout this
parameter space.  As such, we will compare this to the completeness
measured for HIPASS.

\begin{figure}
\begin{center}
\includegraphics[scale=0.7, angle=-90]{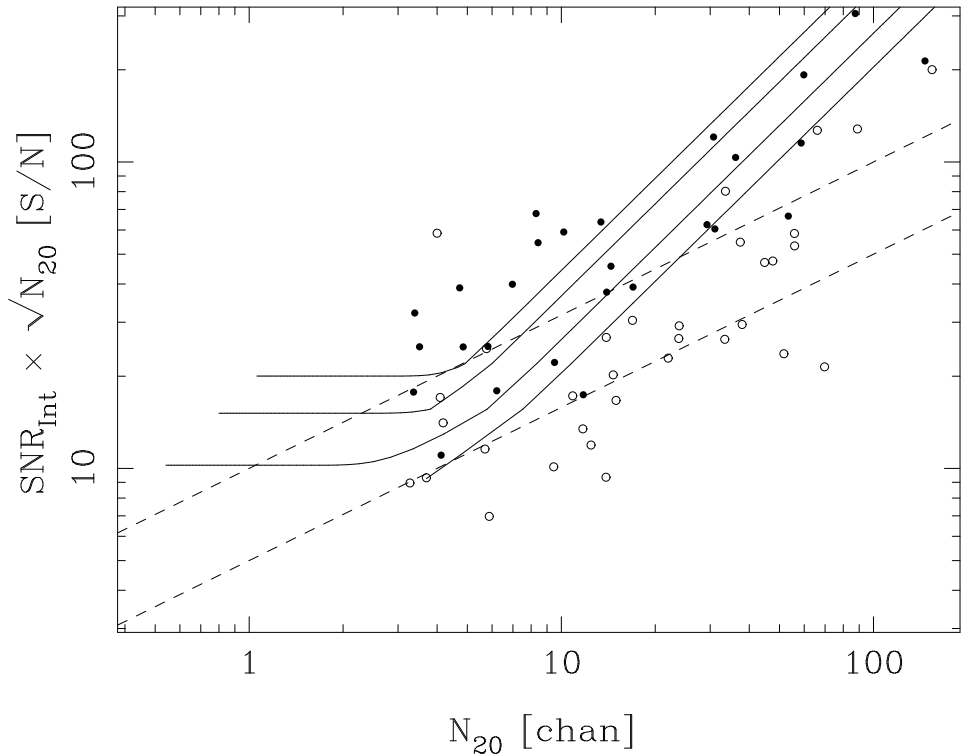}
\caption{The completeness for our survey is shown by scaling the 
HIPASS completeness function from \citet{zwa04} to our channel size and 
sensitivity.  The abscissa is in units of channels and the ordinate is in 
units of signal-to-noise in a single channel.  The solid lines represent 
(from bottom to top) 50\%, 75\%, 
95\%, and 99\% completeness.  The simulated sources from 
Figure~\ref{fig:fakes} 
are plotted here for comparison.  The dashed lines mark the theoretical 
5$\sigma$ (lower line) and 10$\sigma$ (upper line) detection limits.  
\label{fig:hicomp}}
\end{center}
\end{figure}

\citet{zwa04} used large numbers of simulated sources inserted into the
HIPASS cubes to characterize the completeness of that survey and
parameterize it as a function of $S_{int}$, $S_{peak}$, and $W_{20}$.
Again, we scale both $S_{int}$ and $W_{20}$ by the HIPASS noise and
channel size; the result is shown in Figure~\ref{fig:hicomp}.  The simulated 
sources inserted into our data are also shown to illustrate that these two 
measures of completeness are roughly consistent.  Note that the completeness 
is not proportional to $\sqrt{W}$, but degrades more rapidly than that.  For
our analysis, we will use the HIPASS completeness function (as scaled
for each group),  but assume 100\% incompleteness for sources with
$SNR_{int} < 5\sigma$.  While this appears to be reasonable, the reader should
remember that HIPASS had different observing parameters (e.g. channel 
size and sensitivity) and observing technique (pure Declination scans vs. 
basket-weave) and a different source-finding approach (automated vs. manual).
This completeness function may, therefore, not be entirely accurate for 
our survey.  Nevertheless, it is probably a reasonable approximation and what
we will use for the remainder of this paper.  

For reference, our 5$\sigma$~\mhi\ detection limit for a source with a 
velocity width of 30 \kms\ ranges from 0.5--2$\times~10^7$\msun\ (depending
on the group).  The lowest mass detection in any of our groups has 
\mhi$= 1\times 10^7$\msun.  At these level, our survey is quite incomplete.

\section{Survey Results}
\label{results}

A total of 111 galaxies were detected by the combined Parkes/ATCA survey.
Of these, 63 are in the groups while the remaining 48 are background 
galaxies.  All of the galaxies in the optically-selected \citet{gar93} group 
catalog were detected in \HI\ as were all galaxies in the HICAT catalog 
\citep{mey04}.  These 63 galaxies represents a doubling of the number of 
galaxies in the optically-selected groups and a 60\% increase over the number 
of HICAT galaxies in these groups.  

All the detected galaxies have optical counterparts, either cataloged in NED or
visible in the Digital Sky Survey.  There were no intergalactic \HI\ clouds 
without optical counterparts detected--no HVC analogs detected.  The 
properties of the newly detected group galaxies are consistent with them being 
late-type spiral, irregular, and dwarf irregular galaxies; only one background 
galaxy is classified as an early-type and is relatively gas-poor.  The 
properties of all our detections and the ensemble properties of the groups 
will be presented and discussed in detail in \citetalias{pis07}.

\section{Discussion}
\label{disc}

\citetalias{pis04} contained the limits on the distances to the
compact high-velocity clouds (CHVCs) based on our non-detection of any
analogs in three of the loose groups we surveyed and a simple model
for the distribution of CHVCs.   In this paper we will use our improved
characterization of the completeness of our survey, in concert with our
model from \citetalias{pis04}, to place limits on the CHVC distribution 
for our entire sample of six loose groups.  Furthermore, we will compare
the limits for both a Gaussian distance distribution and a 
\citet[NFW;][]{nav96,nav97} distribution of CHVCs, and examine the robustness 
of our models in light of their potential weaknesses.  

\subsection{A model for CHVCs}

Our simple model was originally described in \citetalias{pis04}.  That
description is presented again here.  This model basically assumes
that the distribution of CHVCs around the Milky Way is representative of
the distribution around other galaxies.  As such the non-detection of
CHVCs within our sample of six groups provides a  constraint on their
distribution around the Milky Way.  There are 270 CHVCs in  the
catalogs of Milky Way HVCs of \citet{put02} and \citet{deh02a} with
measured fluxes and velocity widths.  We assume they are distributed
with a three-dimensional Gaussian distance distribution centered on the
Milky Way with a given D$_{HWHM}$.  Given this, we ask
for what D$_{HWHM}$'s would we expect zero detections of analogs in
our sample of six groups.  This is done as a Monte Carlo simulation
with 10,000 trials for a range of D$_{HWHM}$ between 40-300 kpc and a
population of CHVCs ranging from 27 to 1728 clouds (0.1-6.4 times the
number of cataloged Galactic CHVCs).  This range in the number of 
CHVCs per group reflects possible variations of the population with
group mass.  If CHVCs are associated with dark matter halos, then we
expect the total number of CHVCs to scale in proportion to the group
mass \citep{kly99}.  The dynamical masses of these groups are within a 
dex of the Local Group \citepalias{pis07}, so this range should be
sufficient.  

We have refined the model for this paper.  Previously, a CHVC was considered 
to be detectable if it was above the theoretical 10$\sigma$ sensitivity for
its velocity width.  Now, we use the full completeness function as
presented in Figure~\ref{fig:hicomp} with a cutoff at 5$\sigma$ for
each group, acknowledging that we can detect
sources below our previous 10$\sigma$ detection limit, but at a
significantly decreased level of completeness.  Effectively, this
provides stronger constraints on the distance and mass limits of the
CHVCs.  The caveats of this model are the same as discussed in
\citetalias{pis04}, and fundamentally assume that the properties of Milky
Way CHVCs are representative of those around other galaxies and that our 
assumed distance distribution is a reasonable approximation of the real
distribution.  The former caveat is the most important as our survey will 
not detect the vast majority of CHVCs, but only the highest mass ones.  
This is discussed more below.  

We have also tested a more physically-motivated distribution of CHVCs by
utilizing a NFW density distribution: 
$\rho\propto\frac{1}{(r/r_s)(1+(r/r_s)^2)}$, where $r_s = r_{vir}/C$ with
C being the concentration.  We vary the value of $r_s$ between 15 and 35
kpc which corresponds to the rough ranges of C and $r_{vir}$ for the Milky
Way \citep{kly02}.  We cutoff the distribution of CHVCs at 10$\times\ r_s$ 
(150-350 kpc) spanning the \citet{kly02} value of $r_{vir} = 258 kpc$.  
Our choice for this cutoff assumes that all the CHVCs are within the virial 
radius of the Milky Way.  

\subsection{Limits on the distances to CHVCs.}

\begin{figure}
\begin{center}
\includegraphics[scale=0.65, angle=-90]{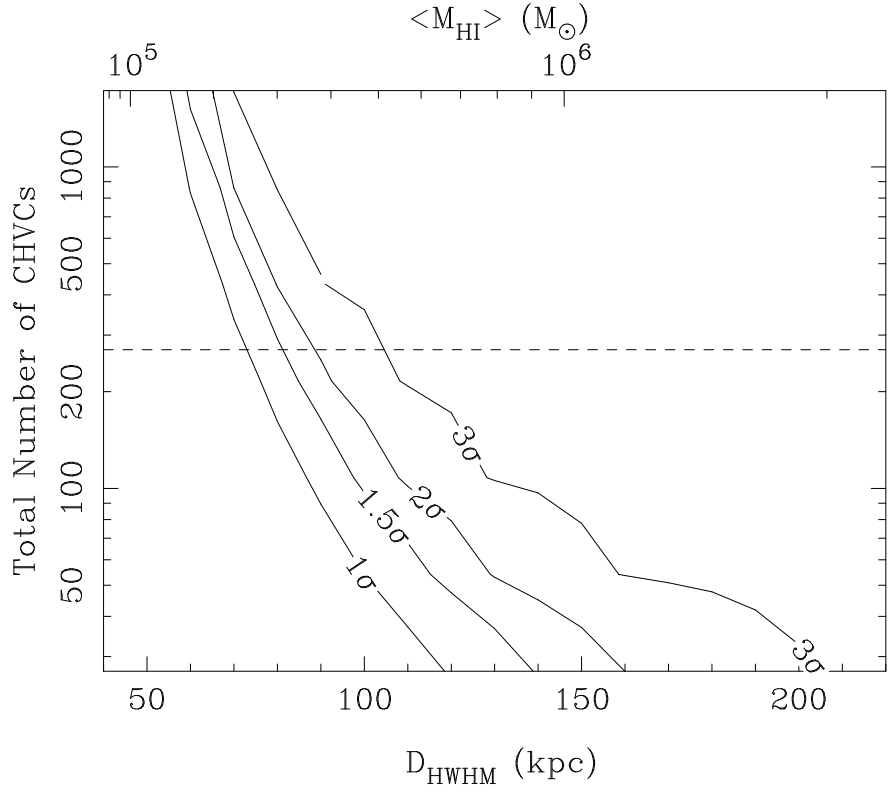}
\caption{Combined constraints on the Milky Way CHVC population as a function 
of the number of CHVCs per group and D$_{HWHM}$ (or the average \HI\ mass
of the CHVC) assuming a three-dimensional Gaussian distance distribution.
The dashed line marks the number of CHVCs identified around the Milky Way.
\label{fig:gausdist}}
\end{center}
\end{figure}

\begin{figure}
\begin{center}
\includegraphics[scale=0.65, angle=-90]{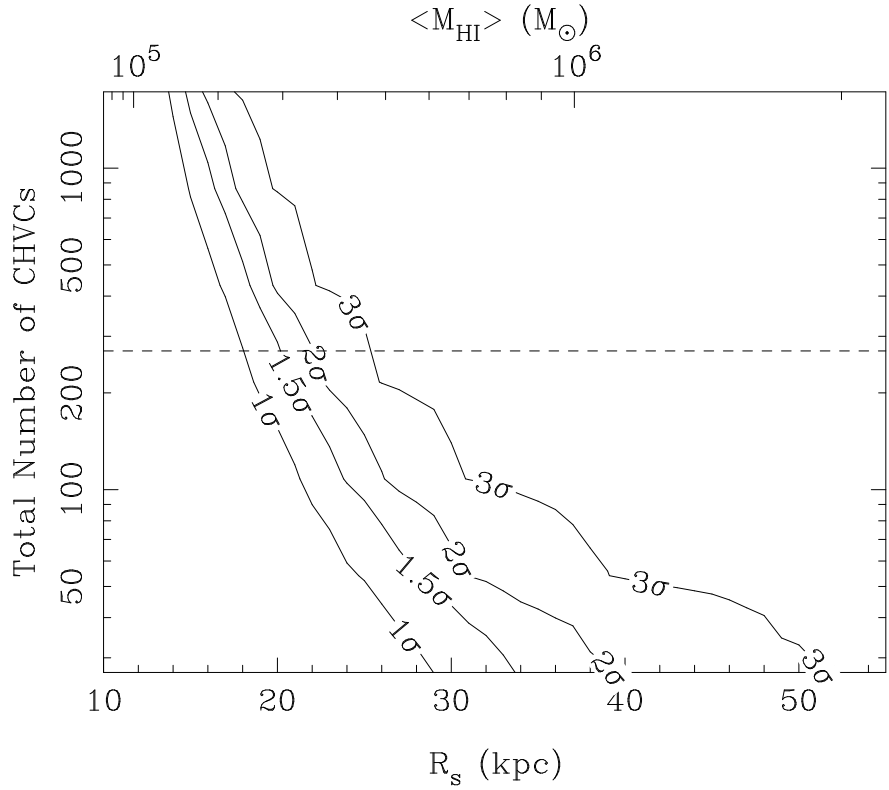}
\caption{As in Figure~\ref{fig:gausdist}, but plotted as a function of $r_s$ 
for an NFW distribution.\label{fig:nfwdist}}
\end{center}
\end{figure}

\begin{figure}
\begin{center}
\includegraphics[scale=0.75, angle=-90]{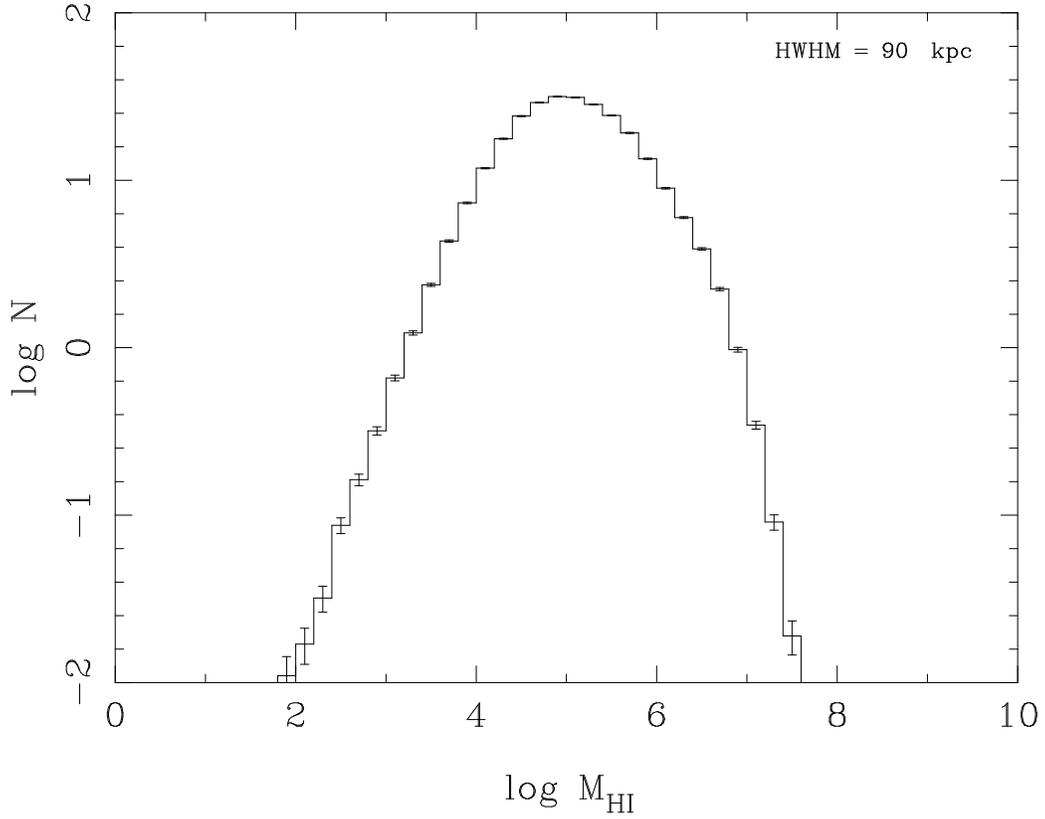}
\caption{The \HI\ mass function for Milky Way CHVCs with a Gaussian 
distance distribution with D$_{HWHM} = 90 kpc$.  The error bars reflect
the standard deviation in each mass bin.\label{fig:mass}}
\end{center}
\end{figure}

The results of modeling the combined sample of six loose
groups are shown in Figures~\ref{fig:gausdist} and \ref{fig:nfwdist}.  
Figure~\ref{fig:gausdist} shows the combined constraints on the number of 
CHVCs per group and D$_{HWHM}$ of the Gaussian distance distribution as
derived from the non-detection of any CHVC analogs in the six groups.
Each D$_{HWHM}$ corresponds to an average \mhi\ for the CHVCs.  The figure
shows that at the 95\%, 2$\sigma$, confidence level a Milky Way-like 
population of 270 CHVCs should be distributed with a D$_{HWHM}\lesssim$90 kpc 
implying an average \mhi\ for CHVCs of $\lesssim\ 4 \times\ 10^5$\msun\ with 
the total population having  \mhi$\lesssim 1\times 10^8$\msun.  The median 
\mhi\ is less than $1.0 \times\ 10^5$\msun\ and the median CHVC distance is
$\lesssim$ 116 kpc.  Compared to our prior limit of D$_{HWHM}\lesssim$160 
kpc \citepalias{pis04}, these limits imply a more tightly clustered, less 
massive population of CHVCs than previously inferred\footnote{In 
\citetalias{pis04}, our derived \mhi\ limits were incorrect.  For 
\citetalias{pis04}'s D$_{HWHM}$ limit of 160 kpc, the average \mhi\ 
should be $\lesssim 10^6$\msun.}.  The CHVC \HI\ mass
function (\HI~MF) for D$_{HWHM} = 90$kpc is shown in Figure~\ref{fig:mass}.  
The shape of the \HI~MF is independent of distance, although the mean \mhi\
will shift with distance.  This \HI~MF is calculated statistically for a 
population of 270 clouds.  It is evident from this figure that, while we do
not expect to detect the average CHVC, we expect approximately 10 CHVCs to 
have \mhi$\ge 5\times 10^6$\msun.  Our survey is sensitive to such clouds, but
is highly incomplete for these masses.  The improvement in our limits as
compared to \citetalias{pis04} comes from the refinement of our completeness 
criteria and the addition of three additional groups to our sample, although 
the former has a larger effect than the latter.  The change in the estimates
of the group distances had no net effect.  

Figure~\ref{fig:nfwdist} shows the limits for an NFW halo density distribution 
as a function of $r_s$.  At the 95\% confidence level, with this distribution, 
the model suggests that $r_s < 22$ kpc for a Milky Way-like population of 270 
CHVCs.  This corresponds to an average \mhi\ for CHVCs of $\lesssim\ 3.4 
\times\ 10^5$\msun\ with the total population having  \mhi$\lesssim 9\times 
10^7$\msun. In this model, the median \mhi\ for CHVCs is $\lesssim\ 4 \times\ 
10^4$\msun\ and the median distance to a CHVC is $\lesssim$ 90 kpc.  These 
limits are very similar to the Gaussian distribution and reflect the 
uncertainties from assuming different distance distributions.  It is worth 
noting that a value of $r_s = 22$ kpc corresponds to the best fit model 
parameters for the Milky Way halo from \citet{kly02} with $r_{vir} = 258$ kpc 
and $C = 12$.  

Our model, while simple, appears to be fairly robust when considering
the possible unique nature of the Milky Way CHVC population.  If we
ignore the most easily detectable CHVC analogs, our limits are
loosened but not dramatically.  Ignoring the two most easily detectable
CHVCs in each simulation, our distance limits slip to match those in 
\citetalias{pis04}. 
The D$_{HWHM}$ limits are not strongly dependent on the total population 
of CHVCs; the limit only decreases by a factor of two when the population 
increases by a factor of 64.  As a result, if we were to consider all 
HVCs, not just CHVCs, our limit should only decrease to 60-70 kpc.  
Finally, we can assume that the \HI\ fluxes of the 270 CHVCs follow a power law
with a slope of $-2.1$ over a range of $4-1000 Jy$\kms\ as described in 
\citet{put02} and their FWHM velocity widths are described by a Gaussian with 
a mean of 36 \kms\ and a dispersion of 12 \kms as described by \citet{deh02a}.
The flux limits span the full range from the minimum to the maximum 
cloud flux in the \citet{put02} catalog.  This removes any effects of the 
discreteness in the cataloged properties on the model results.  In the end, 
however, the resulting distances and masses are the same with 
$D_{HWHM} < 100$ kpc.

For the NFW distribution, there is a stronger dependence on the total 
population, but the critical assumption is the choice of cutoff radius; in
this case, we have chosen a value similar to the virial radius.  If 
there is no cutoff to the distribution, then our model provide no distance 
constraints.  Neither model places an artificial upper limit on the most 
massive CHVCs, but the mass distribution in both cases truncates around 
$10^7$\msun.  

The limits are not strongly sensitive to the distance estimates
to the groups.  D$_{HWHM}$ varies linearly with the distance to the
group, so that the 20\% uncertainty in the group distance results in an 
uncertainty of 20\% on our derived D$_{HWHM}$ limits.  

\subsection{Comparison of Distance Limits to Past Work}
\label{pastwork}

Our results represent the tightest limits on the distribution of CHVCs
based on observations of galaxy groups, but they are roughly
consistent with the observational and theoretical results of others.
\citet{zwa01} conducted a similar survey and also failed to detect any
HVC analogs with resulting mass and distance limits of $\sim
10^6$\msun\ and $\sim 200$kpc.  Theoretical models of HVCs in dark
matter halos by \citet{mal03,ste02,deh02b} all suggest that the CHVCs
should be within $\sim 150$kpc.  Theoretical models of HVCs as
cooling, condensing clouds in a hot Galactic halo also place HVCs at
distances of $\sim 150$ kpc \citep{mal04}, although detailed
simulations suggest they may be much closer $\sim 10-60$ kpc
\citep{kau06,som06}.  Studies of HVCs around other galaxies have
identified populations of clouds which are often associated with the 
optical disk, which are probably associated with a galactic fountain 
\citep[e.g.][]{swa97,boo05,fra02,bar05}.  Some of these clouds are seen 
beyond the optical disk, but are still within $\sim 50$ kpc of the galaxy.
Other \HI\ clouds that may have a tidal origin are within $\sim 5-100$ kpc 
\citep{sch83,sch85,wil96,hun98,wil98,pis98,pis02,wil04}.  Finally, those
clouds seen around M~31 are all within 50 kpc 
\citep[e.g.][]{thi04,bra04,wes05b,mil05}.  

There are also constraints on the distances to Galactic HVCs from studies of 
the population around the Milky Way.  The only direct constraints come from
observations of absorption lines towards halo stars \citep[e.g.][]{wak01}.  
At present, there are only two complexes with bracketed distances.  Complex A 
is between 8 kpc and 10 kpc \citep{vanW99,wak03}, and complex WB is at 
approximately 8 kpc \citep{tho06}.  However, there are many more distances 
estimates using indirect methods.  Many CHVCs show signatures of ram-pressure 
interaction with an ambient medium, 
such as a head-tail or bow shock morphology \citep{bru00,wes05a}.  
The required density of this medium suggests that CHVCs should have 
distances $\sim 50-150$kpc \citep{qui01,bru01,wes05a}.  If HVCs are close
enough to the Milky Way, then they may be partially ionized by escaping
photons and emit in H$\alpha$.  Detections of H$\alpha$ emission from HVCs 
by \citet{put03, wei03} and \citet{tuf02} place the clouds within the 
Galactic halo at distances of $\sim 40-100$kpc.

\subsection{What are the high velocity clouds?}

Our results are independent of the nature of CHVCs.  They could be 
associated with a galactic fountain, tidal debris, low mass dark matter halos, 
or condensing clouds as long as they are within $\sim$ 90 kpc of the Milky Way,
have average \mhi\ of $\lesssim 4\times 10^5$\msun, and a total \mhi\ of 
$\lesssim 10^8$\msun (considering just the CHVCs).  Given these limits, can 
we discriminate between different possible origins for CHVCs?

\begin{figure}
\begin{center}
\includegraphics[scale=0.70, angle=-90]{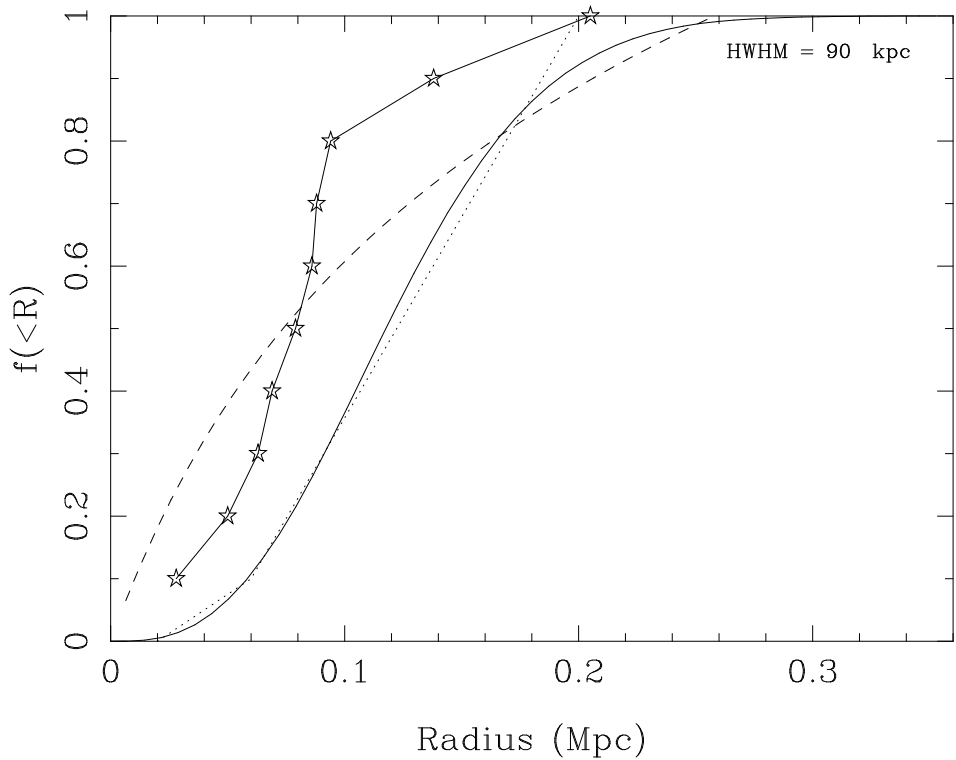}
\caption{The cumulative distance distribution of Milky Way satellites
(solid line with stars), the \citet{kra04} CDM subhalos (dotted line), and 
for a Gaussian distribution with D$_{HWHM} = 90 kpc$ (solid line) and
a NFW distribution with $r_s = 22 kpc$ (dashed line).\label{fig:distances}}
\end{center}
\end{figure}

We have already noted that possible extragalactic analogs to HVCs are all 
observed within $\sim 100$ kpc of the galaxy and completely consistent with 
our distance constraints independent of their proposed origins.  As seen in 
Figure~\ref{fig:distances}, our limits are also consistent 
with the HVCs being distributed like the Milky Way satellites.  The median 
distance to the Milky Way satellites within 200 kpc is 69 kpc \citep{gre03}, 
Our upper limits on the median distance to CHVCs are 116 kpc for a Gaussian 
distribution and 90 kpc for an NFW distribution.  If associated with dark 
matter halos, the Milky Way satellites will trace, possibly with some bias, the
dark matter distribution of the Milky Way.  

\citet{kra04} have identified CDM subhalos in their simulations that could 
accrete gas and form stars.  The distribution of their simulated luminous 
satellites matches that of the Milky Way population of dwarf galaxies.  
The distribution of all dark matter halos has a median distance of 116 kpc
\citep{kra04}.  Furthermore, they have identified those halos that have 
associated gas with masses greater than $10^6$\msun; corresponding
to a \mhi$> 10^5$\msun.  They predict there should be 50-100 such 
gas clouds within 200 kpc, and only 2-5 within 50 kpc.  This is far fewer
than are seen around M~31 \citep{thi04}.  Given that there are $\sim 300$
CHVCs around the Milky Way, this implies that a subset of CHVCs could 
be associated with dark matter halos but the majority are not.  

Can the masses of these systems tell us anything about the origin of HVCs?  
For extragalactic \HI\ clouds 
with a putative galactic fountain origin the individual clouds have
\mhi$\sim 10^{6-7}$\msun, with the total \mhi\ of all such clouds around a 
galaxy $\sim 10^{7-9}$\msun \citep{sch94,kam93,swa97,fra02,boo04,boo05,bar05}.
For tidal debris, they tend to be more massive with \mhi$\sim 10^{6-9}$\msun\
\citep{sch83,sch85,wil96,hun98,wil98,pis98,pis02,wil04}.  For HVC analogs 
associated with dark matter halos, models suggest that these should have 
\mhi$\sim 10^{5.5-7}$\msun\ \citep{ste02,deh02b,mal03}.  Similar mass 
estimates arise from assuming that HVCs condense from the hot gas around 
galaxies \citep{mal04}.  Our limit on the total \mhi\ of CHVCs is 
$\lesssim 10^8$\msun.  If we were to account for all of the HVCs, then
this limit will increase slightly.  \citet{put06} suggest that the total mass 
(ionized$+$neutral) in all HVCs should be $\sim 10^9$\msun\ if the clouds 
are within 150 kpc and $\sim 5\times 10^8$\msun\ if they are within 60 kpc 
and assuming that their neutral fraction is 15\%.  Our limits imply that those
galaxies with observed high velocity gas are probably not analogous to
the Milky Way.  In fact most of the systems referenced above are either 
interacting or undergoing vigorous starbursts.  Our limits of the \HI\ mass
of HVCs can not discriminate between possible formation scenarios.  

\section{Conclusions}
\label{conc}

We have surveyed six groups of galaxies analogous to the Local Group in
\HI\ emission using the Parkes telescope and the ATCA, searching
for counterparts to the HVCs seen around the Milky Way--\HI\ clouds lacking
stellar counterparts.  No such \HI\ clouds were found.  If we assume that a
Milky Way-like population of HVCs is present in each of these groups, we can 
infer an upper limit on their masses and distances.  If HVCs are distributed 
with a three-dimensional Gaussian density distribution around galaxies, then 
the data imply that they must be clustered with $D_{HWHM}\lesssim 90$ kpc, and 
an average \mhi$\lesssim 4\times 10^5$\msun.  While these limits are general, 
we were specifically interested in HVCs potentially associated with low mass 
dark matter halos:  the CHVCs.  The total population of CHVCs can only have 
\mhi$\lesssim 10^8$\msun with median values $D\lesssim 116$ kpc and 
\mhi$\lesssim 10^5$\msun.  This limit is not strongly dependent on choosing a 
Gaussian or NFW distance distribution.  Using the latter, the median distance 
drops to 90 kpc with the median \mhi$\lesssim 4\times 10^4$\msun. As such 
there is not a large reservoir of neutral gas around galaxies, and, therefore, 
any significant reservoir of baryons around galaxies must be mostly ionized.  
This is consistent with evidence from absorption line studies of our own 
Galaxy and others \citep[e.g.][]{tri00a,tri00b,nic03,sem03}.

These limits are stronger than previous limits based on searches for 
extragalactic HVC analogs in groups of galaxies \citep{zwa01,pis04}.  Only
those searches for HVCs associated with the optical disks of galaxies 
\citep[e.g.][]{swa97,boo05,fra02} and the survey of HVCs seen around M~31 
\citep[e.g.][]{thi04} are deeper.  Distances inferred for Milky Way HVCs 
tend to be smaller, $\lesssim 50-100$ kpc.  Unfortunately, our limits do not 
place strong constraints on the possible origin of CHVCs.  They are roughly 
consistent both with models that associate them with dark matter halos 
\citep{deh02b} and with those that propose they are condensing clouds in a 
hot circumgalactic halo \citep{mal04}.  Current and ongoing surveys of groups 
of galaxies, such as AGES \citep{aul06} and GEMS \citep{for06,kil06,kil05} 
along with deeper \HI\ observations of other galaxies, more sophisticated 
modeling, and improved understanding of Milky Way HVCs will help constrain 
the origin and nature of these mysterious objects.

\acknowledgements

The authors wish to thank the staff at Parkes and the ATCA for their 
assistance with observing; Warwick Wilson for his excellent 
work in making the 16 MHz filters for these observations; 
Martin Zwaan for his assistance inserting simulated sources into our data
cubes;  Karen Masters for providing the distances to our 
galaxy groups using her velocity flow model; and Jay Lockman and 
Stuart Wythe for their critical readings and insightful comments on the
paper.  This research was performed in part while D.J.P. held a National 
Research Council Research Associateship Award at the Naval Research 
Laboratory.  Basic research at the Naval Research Laboratory is supported by 
6.1 base funding.  D.J.P. also acknowledges generous support 
from NSF MPS International Distinguished Research Fellowship grant AST 0104439 
and partial support from an ATNF Bolton Fellowship.

\clearpage

\begin{deluxetable}{ccccccc}
\tablecolumns{7}
\tablewidth{0pc}
\tablecaption{Group Properties\label{tab:grpprop}}
\tablehead{\colhead{Group} & \multicolumn{2}{c}{Group Center\tablenotemark{a}} 
& \colhead{V$_\odot$\tablenotemark{b}} & \colhead{Distance\tablenotemark{c}} 
& \colhead{Scale} & \colhead{Num. Galaxies\tablenotemark{a}}\\
\colhead{} & \colhead{$\alpha$ (J2000)} & \colhead{$\delta$ (J2000)} & 
\colhead{} & \colhead{} & \colhead{} & \colhead{} \\
\colhead{} & \colhead{h:m:s} & \colhead{$^\circ :\arcmin\ : \arcsec$} & 
\colhead{\kms} & \colhead{Mpc } & \colhead{1$^\circ$ = X kpc}& \colhead{} }
\startdata
LGG 93       & 03:23:32.1 & -52:13:00 &  883 & 10.9 & 190 & 5 \\
LGG 106      & 03:54:56.9 & -47:43:42 & 1068 & 13.8 & 241 & 6 \\
LGG 180      & 09:43:54.6 & -31:31:10 & 1059 & 14.8 & 258 & 8 \\
LGG 293      & 12:34:12.4 & -07:24:46 & 1016 & 11.1 & 194 & 4 \\
LGG 478      & 23:36:12.5 & -36:58:40 &  686 &  8.6 & 150 & 4 \\
HIPASS Group & 13:07:12.4 & -18:31:52 &  793 &  9.1 & 159 & 4 
\enddata
\tablenotetext{a}{From \citet{gar93}.}
\tablenotetext{b}{Average of individual group galaxies' V$_\odot$ from NED}
\tablenotetext{c}{The distance to the group calculated using the multiattractor
velocity flow model of \citet{mas05} assuming 
H$_0$ = 72 km s$^{-1}$ Mpc$^{-1}$.}
\end{deluxetable}

\begin{deluxetable}{cccccccccccc}
\tabletypesize{\footnotesize}
\tablecolumns{12}
\tablewidth{0pc}
\tablecaption{Parkes Observing Properties\label{tab:obsprop}}
\tablehead{\colhead{Group} & \colhead{Distance\tablenotemark{a}} & 
\multicolumn{2}{c}{Group Area} & \multicolumn{2}{c}{Bandwidth} & 
\colhead{$\nu_0$} & \colhead{Beams} & \colhead{Beamwidth} & 
\multicolumn{3}{c}{Sensitivity\tablenotemark{b}} \\
\colhead{} & \colhead{Mpc} & \colhead{sq. deg.} & \colhead{Mpc$^2$} &
\colhead{MHz} & \colhead{\kms } & \colhead{MHz} & \colhead{} & \colhead{14$\arcmin$= X kpc} &
\colhead{mJy} & \colhead{$10^5$\msun} &\colhead{$10^{16}$\cmsq}}
\startdata
LGG 93       & 10.9 & 30 & 1.1 &  8 & 1700 & 1416 & 7  & 44 & 7.0 & 7.2 & 3.6 \\ 
LGG 106      & 13.8 & 25 & 1.5 & 16 & 3400 & 1414 & 13 & 56 & 5.5 & 7.4 & 2.8 \\
LGG 180      & 14.8 & 25 & 1.7 &  8 & 1700 & 1416 & 7  & 60 & 6.5 & 11  & 3.4 \\
LGG 293      & 11.1 & 22 & 0.8 & 16 & 3400 & 1414 & 13 & 45 & 6.0 & 4.8 & 3.1 \\
LGG 478      &  8.6 & 35 & 0.8 & 16 & 3400 & 1413 & 13 & 35 & 6.5 & 3.5 & 3.4 \\
HIPASS Group &  9.1 & 48 & 1.2 & 16 & 3400 & 1414 & 13 & 37 & 9.0 & 5.2 & 4.6 
\enddata
\tablenotetext{a}{The distance to the group from Table~\ref{tab:grpprop}.}
\tablenotetext{b}{The 1$\sigma$ sensitivity over 3.3 \kms.}
\end{deluxetable}

\clearpage

\end{document}